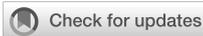

# Deep learning for automatic head and neck lymph node level delineation provides expert-level accuracy

Thomas Weissmann[1,2], Yixing Huang[1,2], Stefan Fischer[1,2], Johannes Roesch[1,2], Sina Mansoorian[1,2], Horacio Ayala Gaona[1,2], Antoniu-Oreste Gostian[2,3], Markus Hecht[1,2], Sebastian Lettmaier[1,2], Lisa Deloch[1,2,4], Benjamin Frey[1,2,4], Udo S. Gaipl[1,2,4], Luitpold Valentin Distel[1,2], Andreas Maier[5], Heinrich Iro[2,3], Sabine Semrau[1,2], Christoph Bert[1,2], Rainer Fietkau[1,2] and Florian Putz[1,2]*

[1]Department of Radiation Oncology, Universitätsklinikum Erlangen, Friedrich-Alexander-Universität Erlangen-Nürnberg, Erlangen, Germany, [2]Comprehensive Cancer Center Erlangen-EMN (CCC ER-EMN), Erlangen, Germany, [3]Department of Otolaryngology, Head and Neck Surgery, Friedrich-Alexander-Universität Erlangen-Nürnberg, Erlangen, Germany, [4]Translational Radiobiology, Department of Radiation Oncology, Friedrich-Alexander-Universität Erlangen-Nürnberg, Universitätsklinikum Erlangen, Erlangen, Germany, [5]Pattern Recognition Lab, Friedrich-Alexander-Universität Erlangen-Nürnberg, Erlangen, Germany

**Background:** Deep learning-based head and neck lymph node level (HN_LNL) autodelineation is of high relevance to radiotherapy research and clinical treatment planning but still underinvestigated in academic literature. In particular, there is no publicly available open-source solution for large-scale autosegmentation of HN_LNL in the research setting.

**Methods:** An expert-delineated cohort of 35 planning CTs was used for training of an nnU-net 3D-fullres/2D-ensemble model for autosegmentation of 20 different HN_LNL. A second cohort acquired at the same institution later in time served as the test set (n = 20). In a completely blinded evaluation, 3 clinical experts rated the quality of deep learning autosegmentations in a head-to-head comparison with expert-created contours. For a subgroup of 10 cases, intraobserver variability was compared to the average deep learning autosegmentation accuracy on the original and recontoured set of expert segmentations. A postprocessing step to adjust craniocaudal boundaries of level autosegmentations to the CT slice plane was introduced and the effect of autocontour consistency with CT slice plane orientation on geometric accuracy and expert rating was investigated.

**Results:** Blinded expert ratings for deep learning segmentations and expert-created contours were not significantly different. Deep learning segmentations with slice plane adjustment were rated numerically higher (mean, 81.0 vs. 79.6, p = 0.185) and deep learning segmentations without slice plane adjustment were rated numerically lower (77.2 vs. 79.6, p = 0.167) than manually drawn contours. In a head-to-head comparison, deep learning segmentations with CT slice plane adjustment were rated significantly better than deep learning contours without slice plane adjustment (81.0 vs. 77.2, p = 0.004). Geometric accuracy of deep learning segmentations was not different from intraobserver variability (mean Dice per level, 0.76 vs. 0.77, p = 0.307). Clinical significance of contour consistency with CT slice plane orientation was not represented by geometric accuracy metrics (volumetric Dice, 0.78 vs. 0.78, p = 0.703).

**Conclusions:** We show that a nnU-net 3D-fullres/2D-ensemble model can be used for highly accurate autodelineation of HN_LNL using only a limited training dataset that is ideally suited for large-scale standardized autodelineation of HN_LNL in the research setting. Geometric accuracy metrics are only an imperfect surrogate for blinded expert rating.

KEYWORDS

deep learning, lymph node level, autosegmentation, head and neck, radiotherapy, neural network, artificial intelligence, target volume

# Introduction

Contouring of head and neck (H&N) lymph node levels is an essential prerequisite for state-of-the-art radiotherapy treatment planning in patients with H&N cancer (1, 2). The draining lymph nodes account for a large portion of the clinical target volume in H&N radiotherapy. Therefore, the role of optimal H&N lymph node level delineation is expected to increase as the use of more conformal radiation techniques like Intensity-modulated radiation therapy (IMRT) and Volumetric modulated arc therapy (VMAT) is becoming more and more widespread (3, 4). Accurate delineation of H&N lymph node levels is of great importance in clinical radiotherapy treatment planning to maximize the chances of a cure while minimizing toxicities in patients with H&N cancer (5, 6). Beyond clinical radiation oncology, precise and standardized delineation of H&N lymph node levels at scale is also especially necessary in radiooncologic research to evaluate and optimize current level definitions and to advance elective treatment of the neck in patients with H&N cancer.

Manual contouring of H&N lymph node levels is a highly laborious and time-consuming task requiring around 50 minutes of an expert's time per patient case (7). These burdening time requirements limit the extent to which optimal H&N lymph node level delineation can be performed in clinical practice, increasing waiting time for patients, limiting frequent adaption of radiotherapy treatment plans and taking away clinical experts from other tasks (8). The large amount of expert time required for manual H&N lymph node level delineation, however, is of particular concern for radiotherapy research, in which lymph node levels need to be delineated at scale and in a standardized manner but with clinical expert time being even more limited.

It is not surprising therefore that there has been great interest in automation of H&N lymph node level segmentation in recent years. However, H&N nodal levels are no simple anatomical structures like organs at risk (OAR) or even primary tumors and lymph node metastases. Instead, H&N nodal levels are complex anatomical compartments defined and separated by a multitude of anatomical landmarks requiring a lot of training time and experience for correct manual delineation by clinical experts (1, 9). Earlier publications focused on atlas-based autosegmentation, which led to some considerable improvements. Overall success was still rather modest however, due to the complexity of H&N lymph node level delineation (10–19). With the advent of deep learning models there have been major advances in autosegmentation for radiation oncology in general and as of now there are even multiple commercial solutions offering deep learning-based H&N lymph node level autosegmentation (20–23). However, despite its commercial availability, deep learning-based H&N lymph node level autosegmentation is still underinvestigated in academic literature. Notably, Cardenas et al. trained a 3D U-net (24) on a dataset of 51 H&N cases to autosegment clinical target volumes composed of multiple instead of individual lymph node levels. The authors found excellent accuracy for nodal clinical target volume (CTV) autodelineation as well as good expert ratings (25). Van der Veen et al. trained a 3D convolutional neural network (3D-CNN) based on the DeepMedic architecture (26) on a clinical dataset of 69 H&N cases to autosegment a selection of 17 individual nodal levels combining level II, III and IV on each side (7). The authors focused on a semiautomatic workflow with a subsequent manual expert-based correction step and were able to show a shortening of manual time required and a reduction in interobserver variability with upfront deep learning autosegmentation. Segmentation accuracy, however, was more modest when no manual correction was used (7).

Despite some ground-laying academic studies and recent commercial availability, there is still no free, high-performing, adaptable open-source solution publicly available that may be used for H&N lymph node level autosegmentation for research purposes. Moreover, to the best of our knowledge deep learning for H&N lymph node level autosegmentation has so far not been compared to manually created expert contours in a fully blinded expert evaluation.

Furthermore, volumetric CT datasets consist of stacks of individual 2D images that are characterized by a certain angulation in reference to the imaged anatomy. Many 3D autosegmentation techniques, e.g., atlas- and 3D-CNN based autosegmentation do not take this CT slice plane orientation into account as they consider 3D volumes instead of individual image slices. However, the definition of head and neck lymph node level boundaries actually relies not only on the location of anatomical landmarks in 3D space but also on the orientation of individual CT slice planes within the 3D image volume (1). We therefore hypothesized that improving consistency of level autosegmentations with the CT slice plane orientation could improve expert assessment of autocontour quality.

## Contributions

In the present manuscript, we proceed with the academic work on H&N lymph node level autosegmentation and make the following contributions. First, we show how the publicly available nnU-Net pipeline (27) can be used for highly accurate autosegmentation of 20 different H&N lymph node levels. We train the nnU-Net on a limited training set of 35 cases and validate the segmentation performance in an independent test cohort using a temporal external validation strategy. Second, in addition to established geometric accuracy metrics we evaluate the quality of deep learning-based lymph node level autosegmentation using a fully blinded evaluation by three experts. Furthermore, we assess the geometric accuracy of deep learning lymph node level segmentations in reference to human expert recontouring accuracy. Third, we show that the consistency of level segmentations with the CT slice plane orientation is not reflected in geometric similarity metrics but is important to expert ratings of contour quality. Fourth, as the CT slice plane geometry commonly is not available to 3D-CNNs, we introduce a simple postprocessing step informed by the network predictions to make autocontours consistent with the CT slice plane.

## Methods

### Patient population, description of cohorts, and temporal external validation strategy

Planning CT datasets included in this analysis were obtained from patients treated with definitive chemoradiation for head and neck squamous cell carcinoma as part of the phase II CheckRad-CD8 trial (NCT03426657) at the Department of Radiation Oncology of the University Hospital Erlangen (28, 29). All patients had provided their written informed consent for their imaging data to be used for further scientific investigations. IRB approval was obtained for this study from the institutional Ethics Committee (approval number 131_18 Az). The phase II CheckRad-CD8 trial was a single arm trial that evaluated the combination of durvalumab, tremelimumab and radiotherapy in locally advanced head and neck cancer patients. Treatment consisted of an induction cycle with cisplatin, docetaxel and durvalumab followed by concurrent radio-immunotherapy in week five (28, 29). Two temporal cohorts of patients from the CheckRad-CD8 trial, for which planning CTs had been acquired and head and neck lymph node levels had been delineated at different moments in time, were used in the present analysis for training and testing, respectively. No selection or exclusion of datasets, e.g., according to nodal size or anatomic similarity, was performed. The first cohort used for training consisted of 35 planning CT volumes acquired between October 2018 and July 2020 on three different Siemens Healthineers CT scanners (SOMATOM go.Open Pro n = 21, Sensation Open n = 13, SOMATOM go.Top n = 1). The independent second cohort used for testing consisted of 20 planning CT volumes acquired between April 2020 and May 2021 and was acquired on a Siemens Healthineers go.Open Pro scanner exclusively. For both cohorts the matrix size was always 512 × 512 and the slice thickness was 3 mm. All patients were immobilized in the radiotherapy treatment position using a thermoplastic mask (Softfix 5-point mask, Unger Medizintechnik GmbH, Mülheim-Kärlich, Germany). Iodine contrast (Imeron 350, Bracco Imaging Deutschland GmbH, Konstanz, Germany) was administered in the absence of contraindications. In total, 33 out of 35 patients in the training set and 17 out of 20 patients in the test set had received contrast (Table 1). The strategy to use the two temporal cohorts for training and testing respectively instead of mixing the datasets and drawing the test set randomly was chosen, because temporal external validation frequently is considered to be the stronger form of validation and a better indicator for model generalizability (30, 31). For example, temporal external validation is recommended by the TRIPOD criteria over random drawing of test set samples (31). As the two temporarily distinct cohorts additionally differed in some aspects

TABLE 1 Characteristics of the training and test data cohort.

| Dataset characteristic | Training cohort (N = 35) 2018 - 2020 | Test cohort (N = 20) 2020 - 2021 | P for difference |
|---|---|---|---|
| Patient age, years | | | 0.700# |
|    Mean (range) | 61.0 (38 – 79) | 59.3 (39 – 76) | |
| Patient sex, n (%) | | | 0.731* |
|    Female | 7 (20%) | 3 (15%) | |
|    Male | 28 (80%) | 17 (85%) | |
| Location primary tumor | | | 0.125* |
|    Oropharynx | 18 (51%) | 16 (80%) | |
|    Hypopharynx | 10 (29%) | 1 (5%) | |

*(Continued)*

TABLE 1 Continued

| Dataset characteristic | Training cohort (N = 35) 2018 - 2020 | Test cohort (N = 20) 2020 - 2021 | P for difference |
| --- | --- | --- | --- |
| Larynx | 5 (14%) | 2 (10%) | |
| Oral cavity | 2 (6%) | 1 (5%) | |
| Primary tumor stage | | | 0.001* |
| T1 | 2 (6%) | 0 (%) | |
| T2 | 14 (40%) | 1 (5%) | |
| T3 | 3 (9%) | 9 (45%) | |
| T4 | 16 (46%) | 10 (50%) | |
| Nodal stage | | | 0.974 |
| N0 | 7 (20%) | 3 (15%) | |
| N1 | 4 (11%) | 2 (10%) | |
| N2 | 18 (51%) | 11 (55%) | |
| N3 | 6 (17%) | 4 (20%) | |
| p16 status | | | 1.000 |
| negative | 24 (69%) | 14 (70%) | |
| positive | 11 (31%) | 6 (30%) | |
| CT scanner model name, n (%) | | | 0.001* |
| SOMATOM go.Open Pro | 21 (60%) | 20 (100%) | |
| Sensation Open | 13 (37%) | 0 (0%) | |
| SOMATOM go.Top | 1 (3%) | 0 (0%) | |
| CT reconstruction filtering kernel, n (%) | | | 0.001* |
| Br40f | 21 (60%) | 20 (100%) | |
| B31s | 13 (37%) | 0 (0%) | |
| Br44f | 1 (3%) | 0 (0%) | |
| IV contrast, n (%) | | | 0.342* |
| Yes | 33 (94%) | 17 (85%) | |
| No | 2 (6%) | 3 (15%) | |
| Matrix, n (%) | | | NA |
| 512 × 512 | 35 (100%) | 35 (100%) | |
| Slice thickness, n (%) | | | NA |
| 3 mm | 35 (100%) | 35 (100%) | |
| Pixel spacing, mm² | | | 0.548# |
| Mean (range) | 1.14² (0.83² - 1.27²) | 1.14² (0.89² - 1.17²) | |

#Wilcoxon rank-sum test, *Fisher's exact test.

including CT scanner model and consequentially CT reconstruction filtering kernel, this strategy was expected to provide a better indication of model generalizability (Table 1). In addition, the temporal external validation strategy was chosen with the specific purpose of this investigation in mind i.e., to facilitate delineation of lymph node levels at large scale in the research setting and in prospective trials, in which patients are recruited and imaging datasets are acquired over time.

## Manual delineation of head and neck lymph node levels

Manual delineation of CT image datasets in both the training and test cohort was carried out by a single expert radiation oncologist (TW) following the guideline by Grégoire et al. (1) and then reviewed by a second expert radiation oncologist (FP). In case of disagreement, a consensus was reached between both experts and the contours were revised accordingly. Lymph node levels in both cohorts were contoured independently from each other at different points in time. Manual lymph node level delineation was performed with syngo.via RT Imagesuite VB60 (Siemens Healthcare GmbH, Erlangen, Germany). A total of 20 lymph node levels were differentiated: level Ia, VIa, VIb and VIIa as well as the bilateral levels Ib, II, III, IVa, IVb, V, VIIb, and VIII. Each of the 20 lymph node levels corresponds to one class that needs to be predicted. Including the background class, which is assigned to all but the 20 nodal levels, a total of 21 classes needed to be predicted by the network. The definition for level V in the present work included Level $V_a$, $V_b$ and $V_c$ according to Grégoire et al. (1), thus the boundary between Level $V_{a+b}$ and $V_c$ just below the transverse cervical vessels was not part of this evaluation. For a subset of n = 10 patients in the independent test set, lymph node levels were recontoured by the same expert after an interval of three months to determine the intraobserver variability.

## Preprocessing

For preprocessing, CT image datasets were first cropped manually using the 3D Slicer (v. 4.10.2) by fitting of a cubic region-of-interest with the frontal skull base, carina, surface of the CT table and the most lateral extent of the humeral heads serving as cranial, caudal, dorsal and lateral landmarks, respectively (32). Subsequently, residual parts of the mask and mask holder were removed by foreground masking using Otsu's thresholding as implemented in 3D Slicer with the following parameters otsuPercentileThreshold 0.01, thresholdCorrectionFactor 0.3, closingSize 9 and ROIAutoDilateSize 2 (Figure 1). While the nnU-net pipeline also performs automatic cropping, this only involves removal of regions of zero intensity (27). The additional semiautomatic preprocessing step was performed to obtain a standardized input volume and a common denominator without institution-specific external equipment, in which the variety of acquired datasets across institutions can easily be transformed to (i.e., a CT dataset from the frontal skull base to the carina, laterally extending to the outer surface of the humeral heads with external equipment and CT table removed, Figure 1). The results obtained for this input volume definition were therefore expected to be representative for other institutions with different imaging protocols, more so than unprocessed datasets. In addition, due to the removal of institution-specific image information irrelevant for the autosegmentation of lymph node levels, generalization was expected to improve. In this regard, the provided model weights can be used more easily by other researchers who wish to apply the trained models on their own data after converting their data to the same input definition. (Model weights available at https://github.com/putzfn/HNLNL_autosegmentation_trained_models). Furthermore, the semiautomatic preprocessing was performed to improve computational efficiency and model performance by removal of irrelevant information.

## Deep learning-based lymph node level autosegmentation

The first cohort (n = 35) was used for model training and optimization using internal cross validation. A nnU-Net ensemble model consisting of a 3D full resolution U-Net and a 2D U-net (27) (Figure 1) provided the best results with respect to internal cross validation in the training cohort and was selected for external validation in the independent test set (second cohort). Both the 3D full resolution and the 2D U-net model were trained in five folds with each fold using 28 datasets for training and 7 for internal cross validation. Each fold was trained for a total of 1000 epochs. The augmentation in the nnU-Net pipeline was adapted for the task of lymph node level segmentation with the other parts of the nnU-Net pipeline being kept unchanged. As lymph node level labels change with mirroring of datasets, mirroring during online augmentation was disabled and training datasets were augmented with mirroring and adaption of label values before starting nnU-Net model training. For all experiments nnU-net version 1.6.6 (27) with Python version 3.7.4, PyTorch version 1.9.0 (33) with CUDA version 11.1 (34) and CUDNN version 8.0.5 was used. Model training, inference and all computations were carried out on a GPU workstation using a Nvidia Quadro RTX 8000 with 48 Gb of GPU memory.

## CT slice adjustment postprocessing

The definition of lymph node level boundaries actually relies not only on the location of anatomical landmarks in 3D space but also on the orientation of individual CT slice planes within the 3D image volume (1). Especially craniocaudal level boundaries (e.g., the boundary between level II and III) are expected to be consistent with the orientation of the CT slice plane. To test the impact of CT slice plane inconsistency on expert ratings and to provide a simple solution to fix slice plane inconsistency with 3D segmentation models for the task of lymph node level segmentations, we propose a simple postprocessing step (Figure 1). In this postprocessing step, specific lymph node levels that per definition are craniocaudally adjacent to each other, e.g., with the level above ending on one CT slice and the level below beginning on the next CT slice (i.e., the interface between both level volumes being parallel to the CT slice plane orientation), are considered mutually exclusive on each CT slice. If predicted label values from mutually exclusive lymph node levels are encountered on the same CT slice, the lymph node level with the highest area on that CT slice compared to the conflicting levels is considered to be the correct prediction. The conflicting label values are then overwritten by this label value that has been predicted for the most voxels by the network. By considering one CT slice after another, the proposed slice plane adjustment postprocessing thus achieves consistency of craniocaudal level boundaries with the CT slice plane. To also achieve consistency of cranial boundaries of topmost levels and caudal boundaries of bottommost levels with the CT slice plane (i.e., where level volumes are craniocaudally neighboring background voxels), the background label value was considered to be the correct prediction for a specific CT slice, if a set threshold of voxels with non-background label values was not reached on a particular CT slice. Specifically, CT slices with 10 or fewer voxels with non-background label values or with non-background voxels decreasing from one slice to another by 80% or more and the next slice containing no voxels with non-background label values were replaced by the background label value. A Python script for this slice adjustment postprocessing is provided as Supplemental File 1. We also share the trained models and the supplementary program code for use in other research projects for download at https://github.com/putzfn/HNLNL_autosegmentation_trained_models.

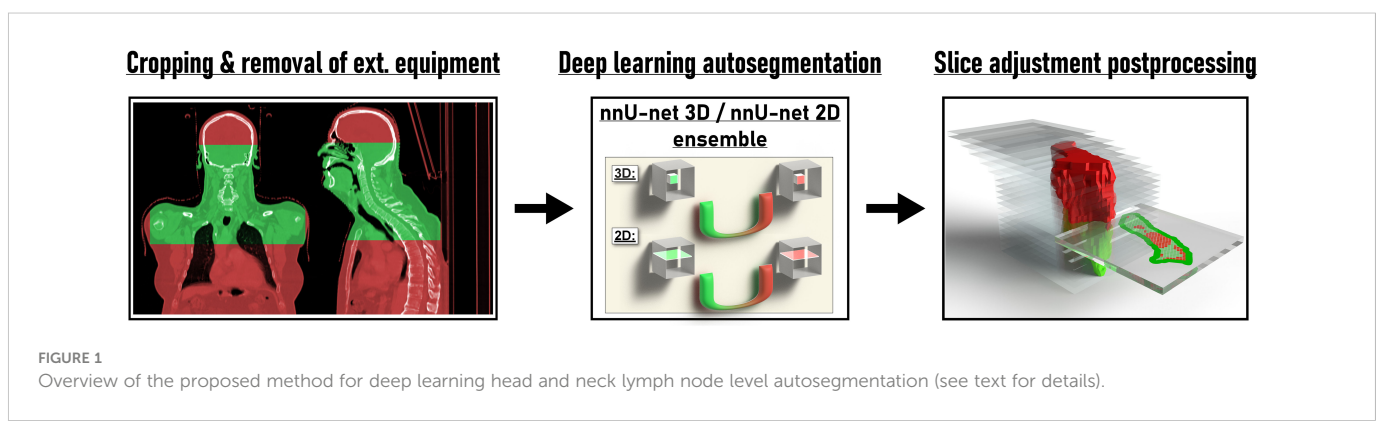

FIGURE 1
Overview of the proposed method for deep learning head and neck lymph node level autosegmentation (see text for details).

## Blinded expert rating

For the independent test set (second cohort, n = 20) three sets of lymph node level contours were evaluated: the expert-generated level contours, the deep learning-generated level contours without slice plane adjustment and the deep learning-generated level contours with the slice plane adjustment postprocessing.

Three independent radiation oncologists who had not taken part in initial contour creation were asked to rate every lymph node level segmentation in the test set on a continuous scale from 0 to 100, guided by four categories (0 – 25: complete recontouring of segmentation necessary, 26 – 50: major manual editing necessary, 51 – 75: minor manual editing necessary, > 75: segmentation clinically usable). The use of a continuous 100-point scale was based on our previous experience that a 4-point Likert scale was insufficient to assess subtle differences in expert judgement of segmentation quality and has similarly been employed by other researchers evaluating expert rating of deep learning predictions (35, 36). The raters were instructed that they would be presented with 60 planning CT datasets with H&N lymph node level segmentations which had been created by human experts or by a deep learning autosegmentation model. The raters were asked to give their judgement on the quality of every lymph node level segmentation but were not informed to look for particular features and were also unaware that they would be presented with two types of autocontour sets (i.e., with and without slice adjustment postprocessing). In total 3600 blinded expert ratings (3 experts × 20 test cases × 20 levels × 3 contour sets) were acquired.

To allow a completely blinded assessment, a dedicated Python module for the open-source software 3D Slicer (v. 4.10.2) (32) was created for this evaluation. In the 3D Slicer module, the experts were presented with one patient case and contour set at a time in a fully blinded way and in randomized order. In the module view the respective lymph node level segmentations were overlayed as contours over the CT image dataset and the experts had the freedom to scroll through all CT slices and also evaluate the contours in coronal as well as sagittal orientation. A digital horizontal slider with a continuous scale of 0 to 100, similar to a visual analogue scale, was used to obtain the expert rating for every level segmentation. The Python code for the blinded expert review module is provided as Supplemental File 2 and at https://github.com/putzfn/HNLNL_autosegmentation_trained_models.

## Evaluation and statistical analysis

Volumetric and surface Dice scores were automatically calculated between individual deep learning-generated level segmentations and the expert-provided ground truth for all test set cases using the Python library surface-distance version 0.1 (37). The tolerance distance for calculating the surface Dice score was set to one voxel. Volumetric and surface Dice score were also calculated considering all levels as a union (i.e., no differentiation between individual levels) to give an estimate for a typical CTV and allow for comparison with previously published work. The maximum Hausdorff distance was calculated using SimpleITK (38).

For the three contour sets, mean volumetric and surface Dice scores as well as mean blinded expert ratings for the 20 test set cases were tested for difference using a paired Wilcoxon signed-rank test. For a subset of 10 patients, the intraobserver variability measured by the volumetric Dice score was calculated after recontouring by the same clinical expert. The intraobserver variability was compared in a paired fashion to the average volumetric Dice score achieved by the deep learning model with slice adjustment postprocessing on the original and recontoured set of expert contours using a paired Wilcoxon signed-rank test. In addition, to compare consistency between manual recontouring and deep learning autosegmentation, the variance in Dice scores between levels was compared in a paired fashion using a paired Levene's test/extended Brown-Forsythe test (39, 40). The Wilcoxon rank-sum test was used to compare two independent groups in the evaluation on the impact of contrast administration. Values are given as mean and median with interquartile range (IQR). P values < 0.05 were considered statistically significant. Statistics were calculated with SPSS Statistics (version 21.0.0.2, IBM) and R (version 3.5.2, R Project for Statistical Computing). Graphs were generated with GraphPad Prism (version 9.5.0, GraphPad Software).

## Results

Deep Learning lymph node level autocontouring took on average 55.6 s per test set dataset on a GPU workstation with slice adjustment postprocessing requiring an additional 0.3 seconds per dataset. Figure 2 shows an exemplary test set patient case with deep learning autocontours and expert-created contours side-by-side. Figure 3 shows the effect of slice adjustment postprocessing.

## Blinded expert rating

For the independent test set (n = 20), three clinical experts provided a total of 3600 blinded level ratings on a scale of 0-100. The mean blinded expert rating was 79.6 (median 80.7, IQR 77.3 – 82.2) for the expert-generated level contours compared to 81.0 (median 81.5, IQR 78.7 – 82.4) and 77.2 (median 76.9, IQR 73.2 – 80.9) for the deep learning-generated segmentations with and without slice plane adjustment, respectively (Figure 4). Blinded expert ratings for deep learning segmentations and expert-created contours were not significantly different. However, compared to the expert-generated contours, deep learning level segmentations without slice plane adjustment had numerically lower ratings (mean difference -2.3, paired p = 0.167), whereas deep learning level autosegmentations with slice plane adjustment were rated numerically higher (mean difference +1.4, paired p = 0.185). Moreover, in a direct comparison, blinded expert ratings were significantly better for deep learning-generated segmentations with than without slice plane adjustment postprocessing (mean difference +3.7, paired p = 0.004). Figure 4 additionally shows blinded expert ratings separately for each of the 20 evaluated head and neck nodal levels. Across all levels, expert- and deep learning-generated level segmentations showed very similar ratings. Slice adjustment postprocessing provided an improvement in blinded expert ratings for most levels. For expert- as well as deep learning-created nodal levels, level Ia obtained the highest expert rating (Figure 4).

## Geometric accuracy

Volumetric and surface Dice score metrics were calculated between deep learning-generated level segmentations and the expert-provided ground truth for the independent test set (n = 20) (Figure 5). Considering all levels as a union, the mean volumetric Dice score was 0.86 (IQR 0.85 – 0.87) with and 0.86 (IQR 0.85 – 0.87) without contours adjusted to the slice plane (median of 0.86 vs. 0.86, paired p = 0.317) with the mean surface Dice score being 0.89 (IQR 0.87 – 0.91) and 0.89 (IQR 0.87 – 0.91), respectively (median of 0.89 vs. 0.89, paired p = 1.00). Considering all lymph node levels individually, the mean volumetric Dice score was 0.78 (IQR 0.76 – 0.79) with and 0.78 (IQR 0.76 – 0.79) without contours adjusted to the slice plane (median of 0.77 vs. 0.77, paired p = 0.703) with the mean surface Dice score being 0.84 (IQR 0.81 – 0.87) and 0.84 (IQR 0.81 – 0.86), respectively (median of 0.84 vs. 0.84, paired p = 0.726). Thus, while blinded expert ratings were significantly better for slice adjustment postprocessing, this difference could not be identified using the volumetric or surface Dice score. Figure 5 additionally shows volumetric and surface Dice score metrics separately for each of the 20 head and neck nodal levels evaluated. Interestingly, in clear contrast to the blinded expert ratings, only a relatively poor volumetric Dice score was achieved for Level Ia (Figure 5). Similar to the observations for volumetric and surface Dice score, no significant difference in maximum Hausdorff distance could be identified between deep learning level contours with and without slice adjustment postprocessing (mean maximum Hausdorff distance, 9.5 vs. 9.3 mm, median 9.3 vs. 9.2 mm, paired p = 0.794).

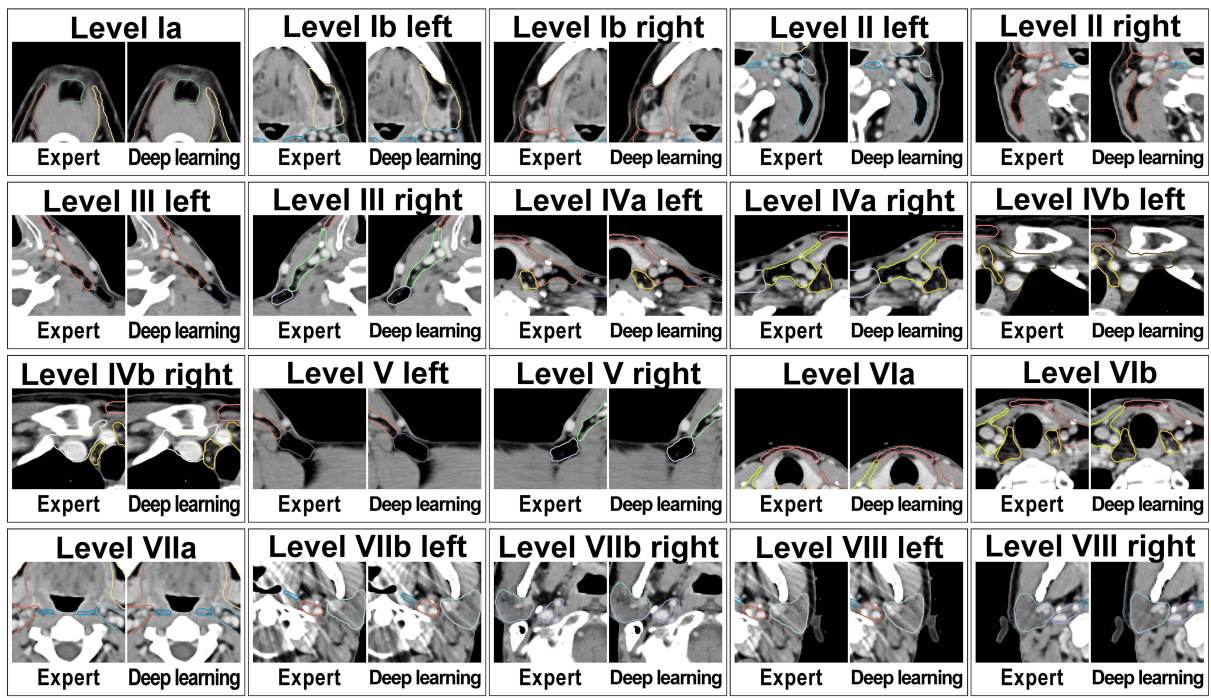

FIGURE 2
Expert-created lymph node level contours and deep learning autocontours side-by-side for each of the 20 evaluated head and neck nodal levels in an exemplary test set patient case.

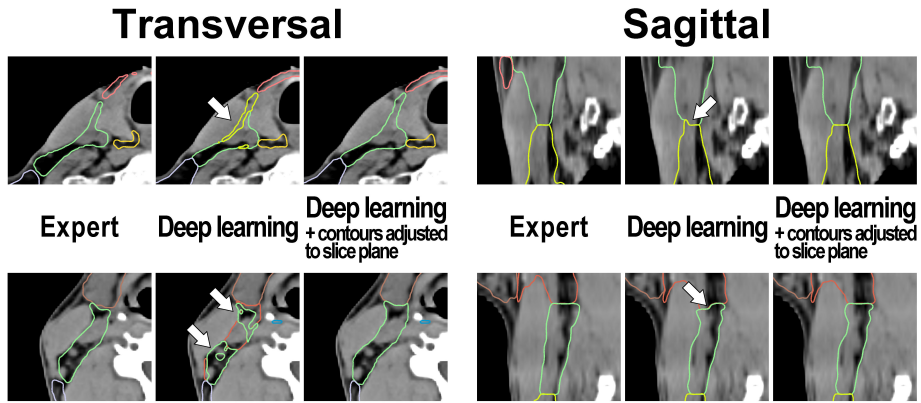

FIGURE 3
Examples illustrating the applied contour postprocessing step that aligns cranial and caudal level boundaries with the planning CT slice plane in test set patient cases. Left: transversal view and Right: sagittal view. Note (arrows): Inconsistencies of level boundaries with the CT slice plane in the output of the deep learning model (center image) that are absent in the expert-generated contour set (left image) and that are removed by the postprocessing step (right image).

## Impact of contrast administration in the test set

Three datasets in the test set were acquired without contrast administration, while the remaining 17 datasets were acquired with contrast. Contrast administration in the test set had no significant impact on geometric accuracy in this study. The mean volumetric Dice score for deep learning autosegmentation (using slice adjustment postprocessing) with contrast administration was 0.77, while it was 0.78 without contrast (median of 0.77 vs. 0.78, p = 0.765). Similarly, no significant impact of contrast administration on blinded expert rating was noted. The mean blinded expert rating for deep learning autosegmentation (using slice adjustment postprocessing) with contrast administration was 80.7, while it was 82.2 without contrast (median of 81.3 vs. 84.4, p = 0.416).

## Comparison to intraobserver variability

Intraobserver variability was assessed for a subgroup of 10 cases in the test set and compared to the average accuracy achieved with the deep learning level segmentations on the original and recontoured set of expert segmentations (n = 10 patients of the independent test set, 200 level segmentations, Figure 6). The mean volumetric Dice score calculated between the original and recontoured nodal levels (= intraobserver variability) was 0.77 (IQR 0.75 – 0.79) and the mean average volumetric Dice score for the deep learning-generated nodal levels (with slice plane adjustment) was 0.76 (IQR 0.75 – 0.78) (median of 0.77 vs. 0.77, mean difference -0.004, paired p = 0.307). The dispersion in the volumetric Dice score for the recontoured level segmentations was larger than the dispersion in the average Dice score for deep learning-generated level segmentations (variance 0.015 vs. 0.011), however a paired robust test for equality of dispersion was not statistically significant (Levene's test/extended Brown-Forsythe test p = 0.219).

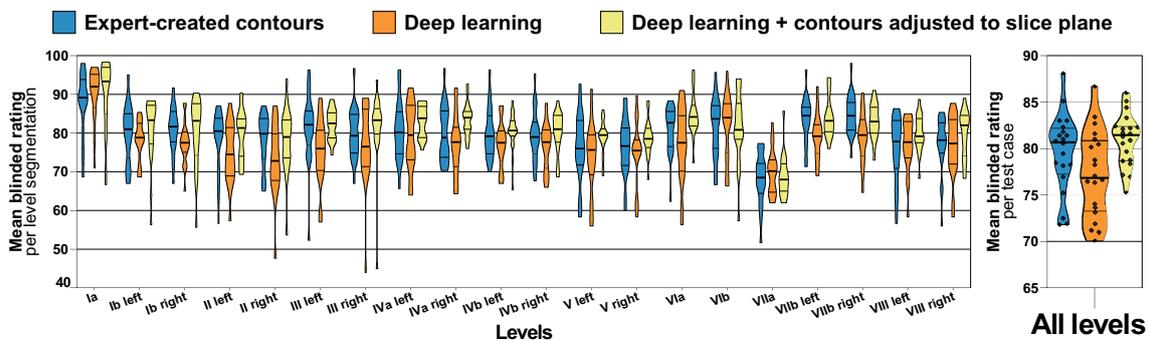

FIGURE 4
Blinded expert rating for expert-created level contours (blue) as well as for deep learning level autocontours without (orange) and with (yellow) the CT slice plane adjustment postprocessing applied (test set n = 20). The left graph shows blinded expert ratings for each level separately and the right graph shows mean ratings per test case across all levels. Improved blinded expert rating with as compared to without slice plane adjustment postprocessing. Violin plots, thick horizontal lines: median, thin horizontal lines: quartiles, rhombi: individual data points.

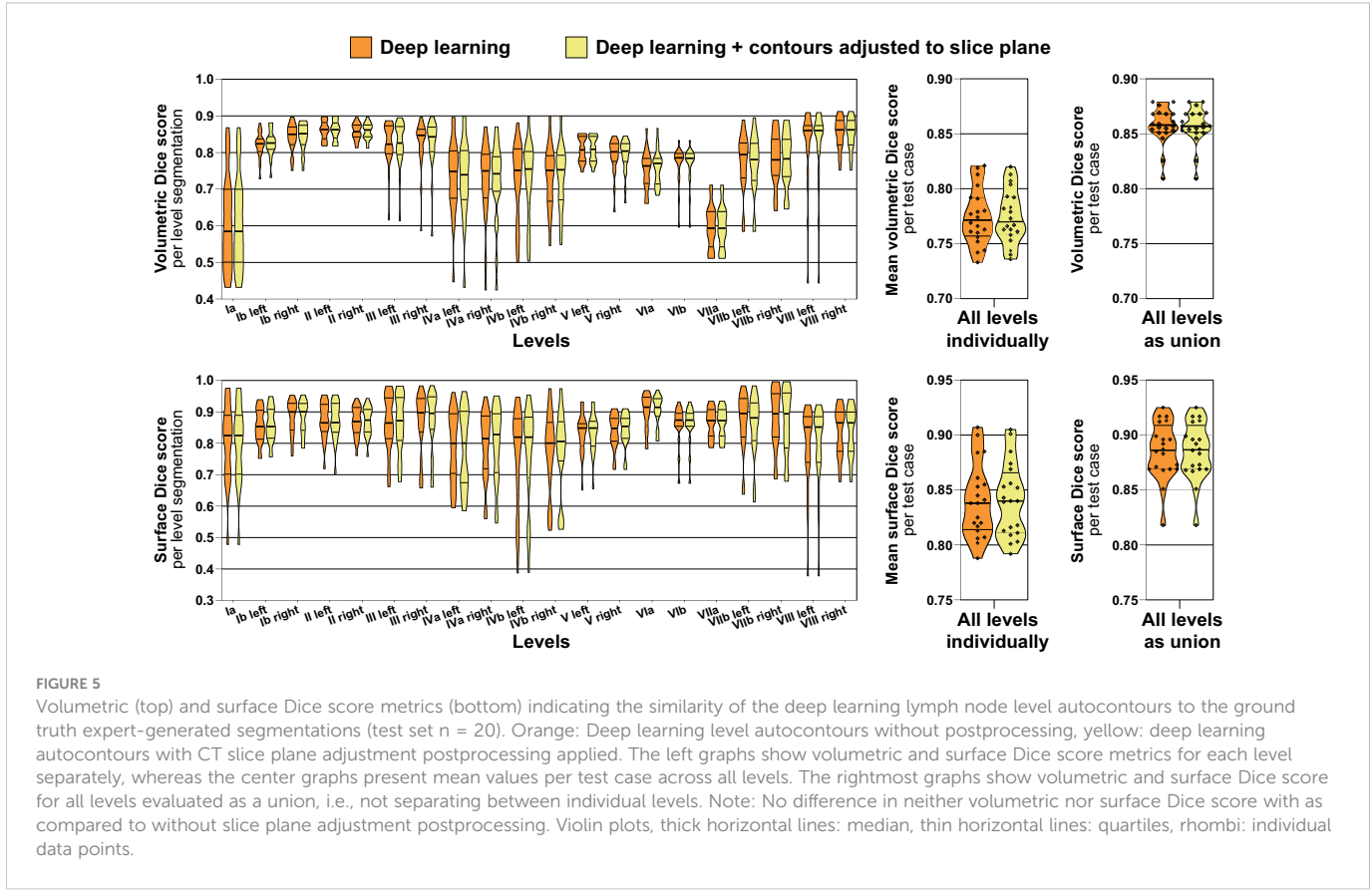

FIGURE 5
Volumetric (top) and surface Dice score metrics (bottom) indicating the similarity of the deep learning lymph node level autocontours to the ground truth expert-generated segmentations (test set n = 20). Orange: Deep learning level autocontours without postprocessing, yellow: deep learning autocontours with CT slice plane adjustment postprocessing applied. The left graphs show volumetric and surface Dice score metrics for each level separately, whereas the center graphs present mean values per test case across all levels. The rightmost graphs show volumetric and surface Dice score for all levels evaluated as a union, i.e., not separating between individual levels. Note: No difference in neither volumetric nor surface Dice score with as compared to without slice plane adjustment postprocessing. Violin plots, thick horizontal lines: median, thin horizontal lines: quartiles, rhombi: individual data points.

# Discussion

## Role of nnU-Net as standardized open-source autosegmentation pipeline

In the present manuscript, we aimed to proceed with the academic work on H&N lymph node level autosegmentation. To allow for optimal reproduction, standardization, and implementation in radiotherapy research we investigated the standardized nnU-Net pipeline for deep learning-based lymph node level segmentation. Unlike commercial products, the nnU-Net pipeline is a publicly and freely available open-source deep learning solution for biomedical segmentation, that is based on open-source Python code and the open-source machine learning framework PyTorch (33). NnU-Net can thus be modified in any way and flexibly integrated in the context of almost any research setting (27). This freedom for modification – not possible with commercial solutions – is essential in the research setting, for example to investigate the effect of novel lymph node level definitions (41). Custom-created U-net implementations are typically highly optimized for individual datasets and thus may contain highly specialized modifications without proven benefit outside the very specific dataset at hand (42). In contrast, the nnU-Net or "no new U-net" automatically adapts to each segmentation task and therefore provides a standardized reference for deep learning-based autosegmentation performance for a variety of segmentation tasks (27). To the best of our knowledge, the present work is the first to show that the nnU-Net with minimum modification and a limited training dataset can be used for highly accurate segmentation of 20 different H&N lymph node levels in an independent test set.

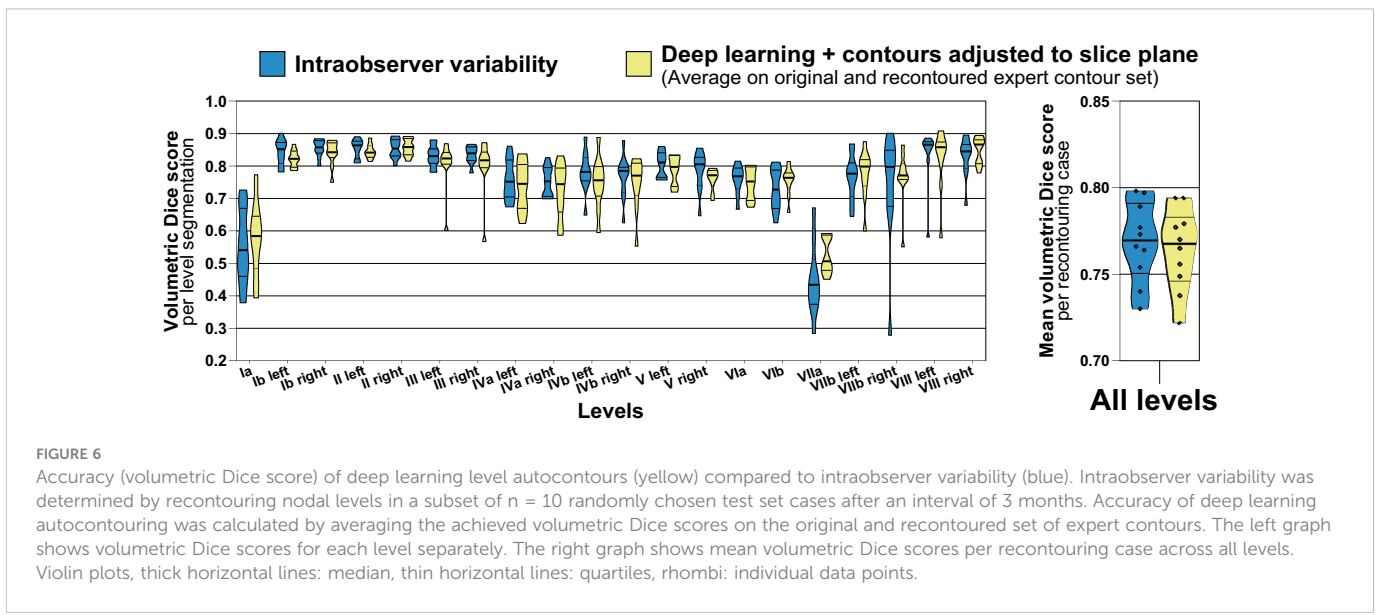

FIGURE 6
Accuracy (volumetric Dice score) of deep learning level autocontours (yellow) compared to intraobserver variability (blue). Intraobserver variability was determined by recontouring nodal levels in a subset of n = 10 randomly chosen test set cases after an interval of 3 months. Accuracy of deep learning autocontouring was calculated by averaging the achieved volumetric Dice scores on the original and recontoured set of expert contours. The left graph shows volumetric Dice scores for each level separately. The right graph shows mean volumetric Dice scores per recontouring case across all levels. Violin plots, thick horizontal lines: median, thin horizontal lines: quartiles, rhombi: individual data points.

## Geometric accuracy of deep learning nodal level autosegmentation

Comparison of Dice scores between different studies is limited due to the high dependency on study design and cohort composition in training and test sets. In addition, intra- and interobserver variability for manual level delineation by experts with a volumetric Dice score on the order of 0.80 and 0.67 is high (7). Cardenas et al. evaluated the segmentation of whole target volumes composed of multiple lymph node levels. This combination of multiple individual nodal levels improves measured segmentation accuracy, as the boundaries between individual levels are irrelevant for the accuracy calculation. Depending on the lymph node CTV, Cardenas et al. reported volumetric Dice scores ranging from 0.81 to 0.90. These values are very comparable to the score of 0.86 observed in our study with all 20 nodal levels considered as a union, especially when taking into account that the selection of nodal levels was different (25). The segmentation accuracy in our study compares favorably to the uncorrected model performance reported by van der Veen et al. (mean volumetric Dice score of 0.78 vs. 0.67 per level). The selection of lymph node levels between both studies was quite similar, but van der Veen et al. considered level II, III and IV as one level, which increases measured geometric accuracy (7).

Because of the limitations regarding direct inter-study comparisons, we used human expert segmentation performance as a reference for comparison. We observed that the accuracy of deep learning-based autosegmentation was not significantly different from intraobserver variability (Dice score of 0.76 vs. 0.77, p = 0.307). Interobserver variability could not be determined, but notably the mean Dice score of 0.78 observed for deep learning autosegmentation in our study was also considerably higher than the mean interobserver variability Dice score of 0.67 measured by van der Veen et al. (value extracted from plot) (7). Intraobserver variability in our study was slightly higher than, but comparable to the investigation by van der Veen et al. (mean Dice score, 0.77 vs. 0.80 [value extracted from plot]), which among other factors can be attributed to the summation of level II, III and IV in the study by van der Veen et al. (7). In the present study, contrast administration in the test set had no effect on autosegmentation performance. This could be explained by the fact that contrast-enhanced and non-contrast-enhanced datasets were part of the training set but this finding could also be partly attributed to the normalization and data augmentation steps in the nnU-Net pipeline. However, the number of patients without contrast in the test set was very low. Further research into the effect of patient- and tumor-related factors on performance of autosegmentation models is warranted and needs to be addressed in future studies with a larger cohort size.

## Role of blinded expert assessment of autosegmentation quality

In a recent critical review, Sherer et al. (43) highlight the importance of physician ratings to assess autosegmentation performance in radiation oncology over geometric similarity metrics. The authors argue that physician assessment of segmentations has been shown to be correlated with clinical outcome whereas geometric measures like the Dice score have not (43). To the best of our knowledge, the present work is the first to have physicians evaluate the quality of deep learning-based lymph node level autosegmentations in a completely blinded head-to-head comparison with expert-created level contours. In this fully blinded evaluation by three radiation oncologists, we observed that deep learning level autosegmentations were not rated significantly different from expert drawn contours. Blinded expert ratings for deep learning-generated level contours with contour adjustment to the CT slice plane were numerically higher than for human expert-created level contours (mean blinded rating, 81.0 vs. 79.6, p = 0.185). Conversely, deep learning-generated level segmentations without slice plane adjustment postprocessing were rated numerically worse (77.2 vs. 79.6, p = 0.167). Moreover, in a head-to-head comparison, deep learning autosegmentations were rated significantly better with than without slice plane adjustment (81.0 vs. 77.2, p = 0.004). Although the magnitude of this effect on expert ratings was more modest and probably would not have been observed using a 4-point Likert scale, contour consistency with the CT slice plane orientation was not reflected in the volumetric nor surface Dice score (mean volumetric Dice score, 0.78 vs. 0.78). This finding may therefore also indicate the importance of integrating clinical experts in the evaluation of autosegmentation models. In addition, acquiring expert ratings on a continuous scale of 0 to 100 could be preferable to lower-resolution point scales to measure small effects on expert preference in regard to autocontour quality.

## Autocontour consistency with the CT slice plane geometry and postprocessing

H&N nodal level definition is actually not only dependent on the location of anatomical landmarks in 3D space but also the orientation of individual CT slice planes within the 3D image volume (1). In the present study we proposed a simple postprocessing step to improve slice plane consistency of 3D-CNN predictions and were able to show a significant improvement in expert ratings. Only CT datasets with a slice thickness of 3 mm were investigated in the present work. However, the observed results should equally apply to datasets with different slice thickness, as the underlying problem is independent from slice resolution (see Methods section for details). The issue of slice plane consistency of 3D-CNNs warrants further research and future studies might investigate more elaborate ways to improve slice plane consistency directly, e.g., by enforcing it in the model itself.

In addition to the slice adjustment postprocessing introduced in the present work, nnU-Net automatically applies a connected component-based postprocessing. This automatic postprocessing step involves removal of all but the largest component for all foreground classes or for specific individual classes, if this empirically improves segmentation accuracy in the cross-validation results (27). Thus, unconnected islands, e.g., spurious dots of isolated level segmentations, that are not connected to the main segmentation of a specific level are automatically removed. Connected component-based postprocessing was also applied in the work by van der Veen et al. with removal of all but the largest component by default for all levels (7). In addition to connected component analysis, Cardenas et al. performed more postprocessing steps that also included "morphology operations such as erosion/dilation with varying filter sizes and dimensionality" as well as removal of holes (25). Individual H&N nodal levels by definition are continuous and singular structures (1). Therefore, connected component analysis and removal of holes can be useful postprocessing steps for lymph node level autosegmentation to incorporate prior knowledge during postprocessing and improve segmentation accuracy. However, extensive postprocessing steps optimized for one particular dataset may generalize poorly and lead to inferior results with external data.

## Required number of training datasets for successful model training

In the present study, a favorable autocontouring performance for deep learning nodal level autosegmentation was achieved with only a small training dataset of 35 annotated planning CT volumes. The number of training data samples was only about half the number used in the previous study by van der Veen (n = 69) and considerably less than employed by Cardenas et al. (n = 51) (7, 25). The favorable autosegmentation performance in our study in the context of a limited training dataset is an important observation and requires additional discussion. It is important to note upfront, however, that the impact of the size of the training set was outside of the scope of this work and of the previous publications by Cardenas and van der Veen (7, 25). Therefore, it is unclear, if the modified 3D U-Net employed by Cardenas et al. would have provided similarly favorable results with less training data. On the other hand, increasing the number of training data samples in this work could have further improved autosegmentation accuracy.

First, there are important considerations in regard to the network architecture and data augmentation pipeline. NnU-Net has already set a new state-of-the-art for most segmentation tasks it was tested on in the past (27). In the present work we employed an ensemble of a 3D and 2D nnU-Net, each trained for five folds, yielding a total of ten model weights. Conversely, only a single 3D CNN with an older architecture (DeepMedic (26)) was used in the work by van der Veen et al. and an ensemble of five 3D U-Net (24) models by Cardenas et al. (7, 25). Moreover, nnU-Net is particularly characterized by an extensive on-the-fly data augmentation pipeline during training, including rotation and scaling, addition of Gaussian noise, Gaussian blurring, change of brightness and contrast, simulation of low resolution, gamma augmentation and mirroring (mirroring was not used on-the-fly in the present work, see Methods section) (27). The relatively low number of samples to yield a favorable model performance for the task of nodal level autosegmentation could thus also be seen as an indication for the efficiency of nnU-Net and its on-the-fly data augmentation pipeline in addition to the strategy of the present work to use an ensemble of a 2D and 3D nnU-net model.

A final consideration is the patient population used for training and testing, respectively. In this analysis, patient datasets with locally advanced head and neck squamous cell carcinoma without any preselection according to nodal stage or anatomic similarity were used. However, all patients in this study were derived from the phase II CheckRad-CD8 trial and had received one course of induction chemotherapy before the planning CT (28, 29). It is therefore to be expected that the disruption of normal anatomy due to lymph node metastases was lower in the present study than in patients without prior induction therapy. Furthermore, all patients underwent definitive chemoradiation and no patient had received prior resection. Prior surgery alters normal head and neck anatomy and may even remove some anatomic landmarks (44). Therefore, manual as well as automatic delineation of head and neck lymph node levels in the postoperative setting is more challenging. Due to the large amount of anatomical variation introduced by surgery, a larger amount of training data is probably required in the postoperative situation.

## Explainability of network predictions

For correct nodal level autosegmentations, the neural network had to learn complex anatomical landmarks or surrogates for these landmarks from the training data itself. The slice adjustment postprocessing was also based on these network predictions, as the number of predicted voxels per slice belonging to one or another class (i.e., level) determined, where the cranio-caudal boundaries were drawn. Explaining neural networks predictions is non-trivial and an important area of current research (45) that was beyond the scope of the present work. However, making autosegmentation networks explainable will be an important focus of our future investigations. Due to the black box character of neural networks, predicted segmentations need to be validated by human experts and corrected if necessary, as the correctness of predictions cannot be guaranteed for every novel input dataset. However, autosegmentation tasks possess an important advantage that facilitates clinical implementation compared to other deep learning tasks, like prognosis prediction (46). While the autosegmentation network's modus operandi may not be transparent, the output of the model - i.e., the autosegmentation - can easily be verified and corrected by human experts. This expert verification of deep learning autosegmentations still is much more time efficient than complete manual delineation and thus may open the way for more complex treatments in radiation oncology.

## Limitations

Limitations of the present study include the fact that only a single set of expert segmentations was available that had been created by consensus of two experts. Consequently, it was not possible to calculate the interobserver variability. It should also be noted that generally a decline in measured segmentation accuracy is expected when employing different experts for segmentation of training and test datasets, because of systematic differences in nodal level segmentation between experts (7). In addition, while the independent test set was significantly different from the training set in multiple aspects and had been acquired in a different period in time (Table 1), training and test set had been obtained from the same institution and generalization of model performance to image datasets from different institutions could not be investigated. We therefore provide the trained models and all employed methods together with this publication for researchers from external institutions for download (see above). As H&N lymph node levels have been defined based on CT imaging data (1), application to other modalities (e.g., PET-CT and MRI) was outside of the scope of the present work. However, the described methods should similarly be applicable to the CT component of PET-CT studies. While the PET volume could be used as a second input channel of the nnU-net, this would not be expected to improve the segmentation accuracy, because of the limited anatomical information provided by PET. Application to MRI datasets needs additional investigation, but is expected to be comparatively more challenging due to the increased variation between individual MRI datasets compared to CT.

## Conclusion

Using the validation strategy employed in this study, we demonstrate that a small number of expert (consensus) level segmentations, created for a limited subset of initially acquired datasets, can be propagated with high accuracy and consistency onto a whole study population using deep learning-based autosegmentation. Geometric accuracy for deep learning autosegmentation was not significantly different from intraobserver variability and blinded expert ratings for deep learning autosegmentations were equivalent to expert-drawn contours. This indicates that for standardized delineation of nodal levels in large cohorts, deep learning autosegmentation may be equivalent to complete manual delineation of all datasets. Moreover, in this use case, expert time can be fully dedicated to creating high-quality delineations for the training subset as well as on review and if necessary, correction of deep learning-generated autosegmentations. We recommend a clinical expert to always manually validate the autosegmented datasets, which is still expected to be more than one order of magnitude faster than manual delineation. The finding that consistency of autocontours with the CT slice plane orientation was not represented by geometric similarity measurements indicates that clinical experts should also be included in model development for radiotherapy autocontouring tasks. On a final note, the authors would like to express their view that target volume autodelineation is an important automation task that may strongly influence the future of the field and should therefore become a focus of academic work in radiation oncology.

## Data availability statement

The datasets presented in this study can be found in online repositories. The names of the repository/repositories and accession number(s) can be found below: https://github.com/putzfn/HNLNL_autosegmentation_trained_models.

## Ethics statement

The studies involving human participants were reviewed and approved by the institutional Ethics Committee of Friedrich-Alexander-Universität Erlangen-Nürnberg (approval number 131_18 Az). The patients/participants provided their written informed consent to participate in this study.

## Author contributions

TW performed the lymph node level contouring, acquisition of raw data and wrote the manuscript draft with FP. FP conceptualized and conducted the technical aspects of this autocontouring study. FP established, adapted, and trained the deep learning model including writing and modification of all necessary program code. FP performed the preprocessing of imaging data and analysis of results including the statistical analysis and creation of figures. YH, SF and AM provided computer science expertise as well as guidance and essential advice for conducting the work. JR, HA, and SM performed the blinded review. MH, SL and SS provided radiotherapy head and neck expertise. LD, BF, UG and LD provided guidance in the conceptualization of the scientific evaluation strategy and provided radiobiologic expertise. CB provided medical physics expertise and crucial scientific guidance. HI and AG provided otolaryngologic expertise. RF provided the resources as well as radiotherapy head and neck expertise and supervision. All authors contributed to the article and approved the submitted version.

## Funding

This work was funded by the Bundesministerium für Bildung und Forschung (BMBF; TOGETHER 02NUK073; GREWIS, 02NUK017G and GREWIS-alpha, 02NUK050E)

## Conflict of interest

The authors declare that the research was conducted in the absence of any commercial or financial relationships that could be construed as a potential conflict of interest.

## Publisher's note

All claims expressed in this article are solely those of the authors and do not necessarily represent those of their affiliated organizations, or those of the publisher, the editors and the reviewers. Any product that may be evaluated in this article, or claim that may be made by its manufacturer, is not guaranteed or endorsed by the publisher.

## Supplementary material

The Supplementary Material for this article can be found online at: https://www.frontiersin.org/articles/10.3389/fonc.2023.1115258/full#supplementary-material

## References

1. Grégoire V, Ang K, Budach W, Grau C, Hamoir M, Langendijk JA, et al. Delineation of the neck node levels for head and neck tumors: A 2013 update. DAHANCA, EORTC, HKNPCSG, NCIC CTG, NCRI, RTOG, TROG consensus guidelines. *Radiother Oncol* (2014) 110(1):172–81. doi: 10.1016/j.radonc.2013.10.010

2. Biau J, Lapeyre M, Troussier I, Budach W, Giralt J, Grau C, et al. Selection of lymph node target volumes for definitive head and neck radiation therapy: A 2019 update. *Radiother Oncol* (2019) 134:1–9. doi: 10.1016/j.radonc.2019.01.018

3. Eisbruch A, Foote RL, O'Sullivan B, Beitler JJ, Vikram B. Intensity-modulated radiation therapy for head and neck cancer: emphasis on the selection and delineation of the targets. *Semin Radiat Oncol* (2002) 12(3):238–49. doi: 10.1053/srao.2002.32435

4. von der Grün J, Rödel C, Semrau S, Balermpas P, Martin D, Fietkau R, et al. Patterns of care analysis for salivary gland cancer: a survey within the German society of radiation oncology (DEGRO) and recommendations for daily practice. *Strahlentherapie und Onkologie Organ der Deutschen Rontgengesellschaft* (2022) 198(2):123–34. doi: 10.1007/s00066-021-01833-x

5. Navran A, Heemsbergen W, Janssen T, Hamming-Vrieze O, Jonker M, Zuur C, et al. The impact of margin reduction on outcome and toxicity in head and neck cancer patients treated with image-guided volumetric modulated arc therapy (VMAT). *Radiother Oncol* (2019) 130:25–31. doi: 10.1016/j.radonc.2018.06.032

6. Mogadas S, Busch CJ, Pflug C, Hanken H, Krüll A, Petersen C, et al. Influence of radiation dose to pharyngeal constrictor muscles on late dysphagia and quality of life in patients with locally advanced oropharyngeal carcinoma. *Strahlentherapie und Onkologie Organ der Deutschen Rontgengesellschaft* (2020) 196(6):522–9. doi: 10.1007/s00066-019-01572-0

7. van der Veen J, Willems S, Bollen H, Maes F, Nuyts S. Deep learning for elective neck delineation: More consistent and time efficient. *Radiother Oncol* (2020) 153:180–8. doi: 10.1016/j.radonc.2020.10.007

8. Olanrewaju A, Court LE, Zhang L, Naidoo K, Burger H, Dalvie S, et al. Clinical acceptability of automated radiation treatment planning for head and neck cancer using the radiation planning assistant. *Pract Radiat Oncol* (2021) 11(3):177–84. doi: 10.1016/j.prro.2020.12.003

9. Rivera S, Petit C, Martin AN, Cacicedo J, Leaman O, Rosselot MCA, et al. Long-term impact on contouring skills achieved by online learning. an ESTRO-FALCON-IAEA study. *Int J Radiat Oncology Biology Phys* (2018) 102(3):e397. doi: 10.1016/j.ijrobp.2018.07.1174

10. Stapleford LJ, Lawson JD, Perkins C, Edelman S, Davis L, McDonald MW, et al. Evaluation of automatic atlas-based lymph node segmentation for head-and-neck cancer. *Int J Radiat Oncol Biol Phys* (2010) 77(3):959–66. doi: 10.1016/j.ijrobp.2009.09.023

11. Gorthi S, Duay V, Houhou N, Cuadra MB, Schick U, Becker M, et al. Segmentation of head and neck lymph node regions for radiotherapy planning using active contour-based atlas registration. *IEEE J Selected Topics Signal Process* (2009) 3(1):135–47. doi: 10.1109/JSTSP.2008.2011104

12. Han X, Hoogeman MS, Levendag PC, Hibbard LS, Teguh DN, Voet P, et al, Wolf TK. Atlas-based auto-segmentation of head and neck CT images. *Med Image Comput Comput Assist Interv.* (2008) 11(Pt 2):434–41. doi: 10.1007/978-3-540-85990-1_52

13. Han X, Hoogeman MS, Levendag PC, Hibbard LS, Teguh DN, Voet P, et al. Atlas-based auto-segmentation of head and neck CT images. *Med image computing computer-assisted intervention MICCAI Int Conf Med Image Computing Computer-Assisted Intervention* (2008) 11(Pt 2):434–41. doi: 10.1007/978-3-540-85990-1_52

14. Chen A, Deeley MA, Niermann KJ, Moretti L, Dawant BM. Combining registration and active shape models for the automatic segmentation of the lymph node regions in head and neck CT images. *Med Phys* (2010) 37(12):6338–46. doi: 10.1118/1.3515459

15. Teng C-C, Shapiro LG, Kalet IJ. Head and neck lymph node region delineation with image registration. *Biomed Eng Online* (2010) 9(1):30. doi: 10.1186/1475-925X-9-30

16. Commowick O, Grégoire V, Malandain G. Atlas-based delineation of lymph node levels in head and neck computed tomography images. *Radiother Oncol* (2008) 87(2):281–9. doi: 10.1016/j.radonc.2008.01.018

17. Daisne JF, Blumhofer A. Atlas-based automatic segmentation of head and neck organs at risk and nodal target volumes: a clinical validation. *Radiat Oncol (London England)* (2013) 8:154. doi: 10.1186/1748-717x-8-154

18. Yang J, Beadle BM, Garden AS, Gunn B, Rosenthal D, Ang K, et al. Auto-segmentation of low-risk clinical target volume for head and neck radiation therapy. *Pract Radiat Oncol* (2014) 4(1):e31–7. doi: 10.1016/j.prro.2013.03.003

19. Haq R, Berry SL, Deasy JO, Hunt M, Veeraraghavan H. Dynamic multiatlas selection-based consensus segmentation of head and neck structures from CT images. *Med Phys* (2019) 46(12):5612–22. doi: 10.1002/mp.13854

20. Samarasinghe G, Jameson M, Vinod S, Field M, Dowling J, Sowmya A, et al. Deep learning for segmentation in radiation therapy planning: a review. *J Med Imaging Radiat Oncol* (2021) 65(5):578–95. doi: 10.1111/1754-9485.13286

21. Limbus AI. (2022). Available at: https://limbus.ai/.

22. Therapanacea. (2022). Available at: https://www.therapanacea.eu/.

23. Kocher M. Artificial intelligence and radiomics for radiation oncology. *Strahlentherapie und Onkologie Organ der Deutschen Rontgengesellschaft* (2020) 196 (10):847. doi: 10.1007/s00066-020-01676-y

24. Ronneberger O, Fischer P, Brox T eds. U-Net: Convolutional networks for biomedical image segmentation. In: *Medical image computing and computer-assisted intervention – MICCAI 2015*. Cham: Springer International Publishing.

25. Cardenas CE, Beadle BM, Garden AS, Skinner HD, Yang J, Rhee DJ, et al. Generating high-quality lymph node clinical target volumes for head and neck cancer radiation therapy using a fully automated deep learning-based approach. *Int J Radiat Oncol Biol Phys* (2021) 109(3):801–12. doi: 10.1016/j.ijrobp.2020.10.005

26. Kamnitsas K, Ledig C, Newcombe VFJ, Simpson JP, Kane AD, Menon DK, et al. Efficient multi-scale 3D CNN with fully connected CRF for accurate brain lesion segmentation. *Med Image Anal* (2017) 36:61–78. doi: 10.1016/j.media.2016.10.004

27. Isensee F, Jaeger PF, Kohl SAA, Petersen J, Maier-Hein KH. nnU-net: a self-configuring method for deep learning-based biomedical image segmentation. *Nat Methods* (2021) 18(2):203–11. doi: 10.1038/s41592-020-01008-z

28. Hecht M, Eckstein M, Rutzner S, von der Grün J, Illmer T, Klautke G, et al. Primary results of the phase II CheckRad-CD8 trial: First-line treatment of locally advanced head and neck squamous cell carcinoma (HNSCC) with double checkpoint blockade and radiotherapy dependent on intratumoral CD8+ T-cell infiltration. *J Clin Oncol* (2021) 39(15_suppl):6007. doi: 10.1200/JCO.2021.39.15_suppl.6007

29. Hecht M, Eckstein M, Rutzner S, von der Grün J, Illmer T, Klautke G, et al. Induction chemoimmunotherapy followed by CD8+ immune cell-based patient selection for chemotherapy-free radioimmunotherapy in locally advanced head and neck cancer. *J Immunother Cancer* (2022) 10(1). doi: 10.1136/jitc-2021-003747

30. Altman DG, Vergouwe Y, Royston P, Moons KG. Prognosis and prognostic research: validating a prognostic model. *BMJ (Clinical Res ed)* (2009) 338:b605. doi: 10.1136/bmj.b605

31. Moons KG, Altman DG, Reitsma JB, Ioannidis JP, Macaskill P, Steyerberg EW, et al. Transparent reporting of a multivariable prediction model for individual prognosis or diagnosis (TRIPOD): explanation and elaboration. *Ann Internal Med* (2015) 162(1):W1–73. doi: 10.7326/m14-0698

32. Fedorov A, Beichel R, Kalpathy-Cramer J, Finet J, Fillion-Robin JC, Pujol S, et al. 3D slicer as an image computing platform for the quantitative imaging network. *Magnetic resonance Imaging* (2012) 30(9):1323–41. doi: 10.1016/j.mri.2012.05.001

33. Paszke A, Gross S, Massa F, Lerer A, Bradbury J, Chanan G, et al. PyTorch: An imperative style, high-performance deep learning library. (2019). doi: 10.48550/ARXIV.1912.01703

34. Luebke D ed. CUDA: Scalable parallel programming for high-performance scientific computing. In: *2008 5th IEEE international symposium on biomedical imaging: From nano to macro*, Paris, France, (2008), pp. 836–8. doi: 10.1109/ISBI.2008.4541126

35. Zhang Q, Burrage MK, Shanmuganathan M, Gonzales RA, Lukaschuk E, Thomas KE, et al. Artificial intelligence for contrast-free MRI: Scar assessment in myocardial infarction using deep learning–based virtual native enhancement. *Circulation* (2022) 146(20):1492–503. doi: 10.1161/CIRCULATIONAHA.122.060137

36. Bart E, Hegdé J. Deep synthesis of realistic medical images: A novel tool in clinical research and training. *Front Neuroinformatics* (2018) 12:82. doi: 10.3389/fninf.2018.00082

37. Nikolov S, Blackwell S, Zverovitch A, Mendes R, Livne M, De Fauw J, et al. Deep learning to achieve clinically applicable segmentation of head and neck anatomy for radiotherapy. *arXiv preprint arXiv:180904430* (2018) 23(7):e26151. doi: 10.2196/26151

38. Lowekamp BC, Chen DT, Ibáñez L, Blezek D. The design of SimpleITK. *Front Neuroinform* (2013) 7:45. doi: 10.3389/fninf.2013.00045

39. Wilcox RR. Comparing the variances of dependent groups. *Psychometrika* (1989) 54(2):305–15. doi: 10.1007/BF02294522

40. Champely S. (2018). Available at: https://CRAN.R-project.org/package=PairedData.

41. Zhao Y, Liao X, Wang Y, Lan W, Ren J, Yang N, et al. Level ib CTV delineation in nasopharyngeal carcinoma based on lymph node distribution and topographic anatomy. *Radiother Oncol* (2022) 172:10–7. doi: 10.1016/j.radonc.2022.04.026

42. Müller S, Weickert J, Graf N. Robustness of brain tumor segmentation. *J Med Imaging (Bellingham)* (2020) 7(6):064006. doi: 10.1117/1.Jmi.7.6.064006.

43. Sherer MV, Lin D, Elguindi S, Duke S, Tan LT, Cacicedo J, et al. Metrics to evaluate the performance of auto-segmentation for radiation treatment planning: A critical review. *Radiother Oncol* (2021) 160:185–91. doi: 10.1016/j.radonc.2021.05.003

44. Robbins KT, Clayman G, Levine PA, Medina J, Sessions R, Shaha A, et al. Neck dissection classification update: revisions proposed by the American head and neck society and the American academy of otolaryngology-head and neck surgery. *Arch Otolaryngol Head Neck Surg* (2002) 128(7):751–8. doi: 10.1001/archotol.128.7.751

45. Saleem R, Yuan B, Kurugollu F, Anjum A, Liu L. Explaining deep neural networks: A survey on the global interpretation methods. *Neurocomputing* (2022) 513:165–80. doi: 10.1016/j.neucom.2022.09.129

46. Bakas S, Reyes M, Jakab A, Bauer S, Rempfler M, Crimi A, et al. Identifying the best machine learning algorithms for brain tumor segmentation, progression assessment, and overall survival prediction in the BRATS challenge. (2018). doi: 10.48550/ARXIV.1811.02629